\documentclass{article}
\usepackage{spconf,amsmath,graphicx,algorithm,algorithmic,adjustbox,setspace}

\usepackage{color}
\usepackage{enumitem}
\usepackage{hyperref}
\usepackage{multirow}
\usepackage{etoolbox,siunitx}
\robustify\bfseries
\usepackage{booktabs}
\usepackage{balance}
\usepackage{amssymb}
\usepackage{bm}
\usepackage{dsfont}
\usepackage{arydshln}
\usepackage{cite}

\usepackage[hang,flushmargin]{footmisc}

\title{Generation or Replication: Auscultating Audio Latent Diffusion Models}
\name{{\shortstack[c]{Dimitrios Bralios$^{1,2}$, Gordon Wichern$^{1}$,
      François G. Germain$^{1}$, Zexu Pan$^{1}$,\\ Sameer Khurana$^{1}$, Chiori Hori$^1$,
      Jonathan Le Roux$^{1}$\thanks{This work was performed while D.~Bralios was an intern at MERL.}}}}
\address{$^1$Mitsubishi Electric Research Laboratories (MERL), Cambridge, MA, USA\\
$^2$University of Illinois Urbana-Champaign, Urbana, IL, USA}

\begin{document}
\ninept
\maketitle
\begin{abstract}
The introduction of audio latent diffusion models possessing the ability to generate realistic sound clips on demand from a text description has the potential to revolutionize how we work with audio. In this work, we make an initial attempt at understanding the inner workings of audio latent diffusion models by investigating how their audio outputs compare with the training data, similar to how a doctor auscultates a patient by listening to the sounds of their organs. Using text-to-audio latent diffusion models trained on the AudioCaps dataset, we systematically analyze memorization behavior as a function of training set size. We also evaluate different retrieval metrics for evidence of training data memorization, finding the similarity between mel spectrograms to be more robust in detecting matches than learned embedding vectors. In the process of analyzing memorization in audio latent diffusion models, we also discover a large amount of duplicated audio clips within the AudioCaps database.
\end{abstract}

\section{Introduction}
\label{sec:intro}

Diffusion models~\cite{song2019generative, ho2020denoising} have quickly become a powerful class of generative models, and are based on an elegant theoretical formulation that only requires a simple denoising network to iteratively generate complex data such as images and sound. Perhaps their most successful application is in training large scale models that generate images from text captions~\cite{ramesh2022hierarchical}. Text-to-image models based on latent diffusion \cite{rombach2022high}, which perform the iterative diffusion process in a low-dimensional latent space, provide a good trade-off in terms of generation quality and computational cost. However, because text-to-image models are so easy to use, thanks to their control by natural language, and produce realistic images, concerns over the intellectual property rights of the data used to train these models has recently begun to emerge. 

While many legal and ethical issues remain open, research attempting to answer the technical question of how diffusion models may memorize and/or copy their training data has begun to appear for the case of images~\cite{carlini2023extracting, somepalli2023diffusion, somepalli2023understanding}. While precisely defining replication in generative models is difficult, evidence that deep networks memorize training data has emerged through work related to the generalization and overfitting of deep neural networks~\cite{feldman2020neural, carlini2022privacy}, along with research on membership inference attacks~\cite{carlini2022membership, tseng2021membership}, which determine whether a particular sample was part of a model’s training set. 
At the same time, we note that another somewhat counterintuitive by-product of that research is duplicate detection, as recent work found duplicated samples in the training set to be more likely to be replicated~\cite{kandpal2022deduplicating}.

\begin{figure}[t!]
    \centering
    \includegraphics[width=\linewidth]{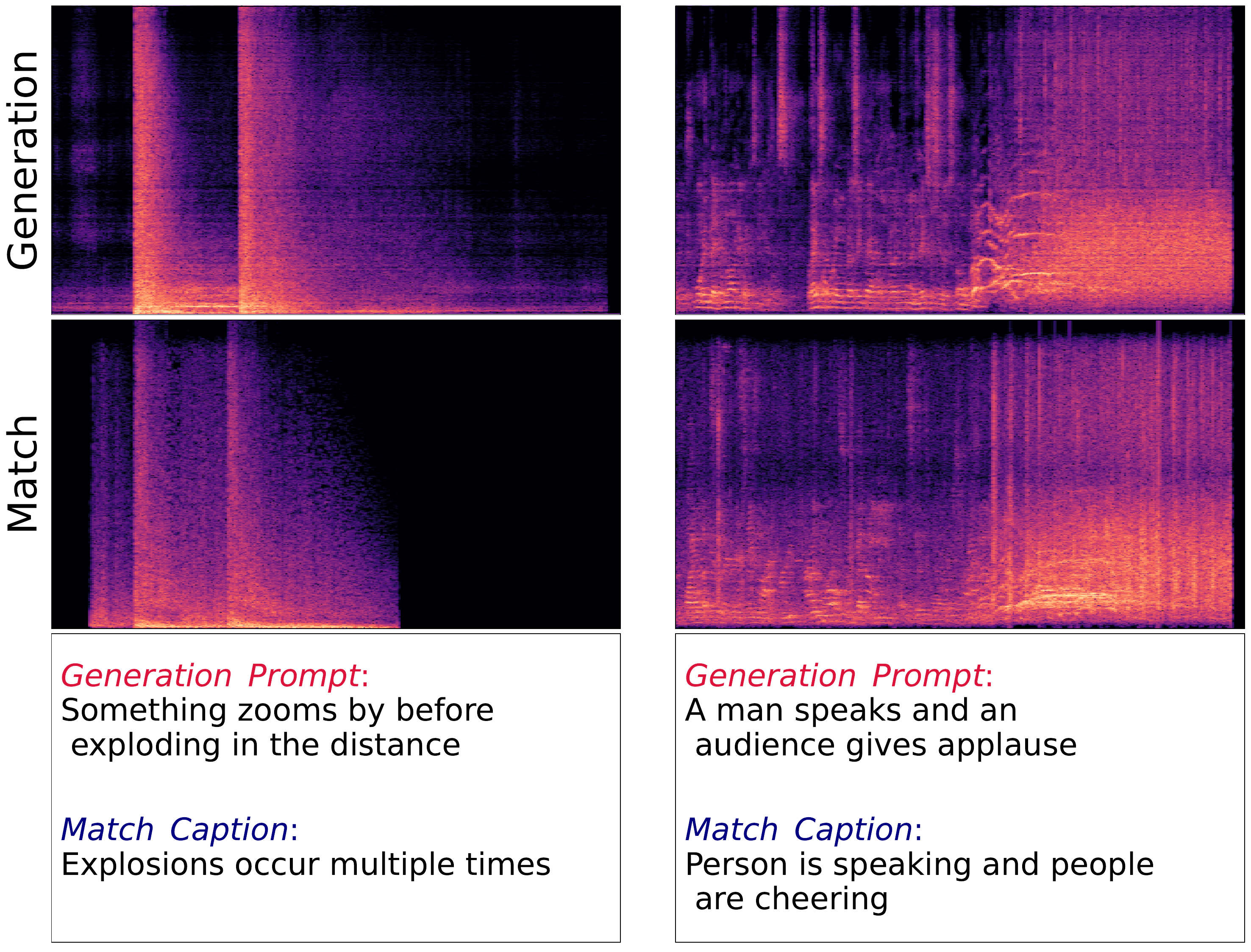}
      \caption{Samples generated using \textsc{Tango}~\cite{ghosal2023tango} and their top matches in the AudioCaps~\cite{kim2019audiocaps} training set along with the corresponding prompts and captions. Found using our proposed method employing mel descriptors. Notice that despite not being identical to the generations, the matches contain semantically identical acoustic events that temporally coincide.}
      \label{fig:fig_one}
\end{figure} 

Diffusion models have also recently found many uses in audio generation for speech~\cite{kong2020diffwave, popov2021grad}, music~\cite{huang2023noise2music, chen2023musicldm}, and general sounds~\cite{yang2023diffsound, liu2023audioldm, huang2023make, ghosal2023tango}. 
The audio latent diffusion model (AudioLDM)~\cite{liu2023audioldm}, which generates mel spectrograms from a text description and then resynthesizes audio waveforms using a separately trained HiFi-GAN network~\cite{kong2020hifi}, is a particularly promising approach.  In a recent public challenge on Foley sound synthesis~\cite{choi2023foley}, a technique based on AudioLDM ranked first in terms of subjective quality. 
While recent works on text-to-music generation have uncovered evidence of training data memorization for transformer-based~\cite{agostinelli2023musiclm} models, and the novelty of generated samples from a diffusion-based model~\cite{chen2023musicldm},  they do not systematically analyze why memorization occurs, or compare the ability of different signal representations to detect replicated training data.

In this work, we make an attempt at quantifying training data replication for audio latent diffusion models trained on general sounds. \textit{We define a generated sound file as replicating training data when it contains nearly-identical complex spectro-temporal patterns.} Certain sounds, such as stationary noise and constant tones, are easily replicated but excluded by this definition since they do not contain such complex patterns. While the term ``nearly-identical'' lacks precision, we believe it is appropriate for this exploratory work, and we propose techniques based on both mel spectrograms and contrastive language audio pre-training (CLAP)~\cite{wu2023large, elizalde2023clap} descriptors for finding these potential replications. Our experiments utilize \textsc{Tango}~\cite{ghosal2023tango}, a text-to-audio model descended from AudioLDM but using a different text encoder and with its diffusion core trained on AudioCaps~\cite{kim2019audiocaps} dataset. We find a small, but not insignificant number of replicated training samples (see Fig.~\ref{fig:fig_one} for examples), and validate our ability to find training data replications by decreasing the size of the training set to a small enough level that memorization is very likely to occur. We also discover a large amount of duplicated audio clips within the AudioCaps database. Audio examples of potentially replicated training data and identified AudioCaps duplicates are available on our demo page~\footnote{\url{https://www.merl.com/demos/auscultating-diffusion}}.

\section{Searching for Replications}
\label{sec:method}

In order to detect generations which potentially are replications of training samples, we frame the problem as a copy detection problem. Specifically, the generated samples are regarded as queries and the training samples as a reference set. For each pair of a query $q$ and a reference $r$, a similarity score is computed, resulting in a pairwise similarity matrix. We can then follow a two stage semi-manual approach consisting of a \textit{retrieval} stage and a \textit{verification} stage. In the retrieval stage, we retrieve the queries $q$ whose top-1 match similarity score is above a certain threshold $\tau$. In the verification stage, we manually examine each retrieved query and reference pair. 

The similarity score $s(q,r)$ between query $q$ and reference $r$ is computed based on a similarity score between descriptors $\mathbf{q},\mathbf{r} \in \mathbb{R}^d$ extracted from $q$ and $r$, in our case their cosine similarity: 
\begin{equation}
\label{eq:sim_score}
   s(q,r)= \text{sim}(\mathbf{q}, \mathbf{r}) = \frac{\mathbf{q}^\top \mathbf{r}}{\|\mathbf{q}\| \|\mathbf{r}\|}.
\end{equation}
We investigate the use of a low-dimensional log mel %
spectrogram as a sample descriptor, a natural choice since most audio models employing latent diffusion extract the latent representation from a mel one. We also experiment with a CLAP~\cite{wu2023large} descriptor.   

Additionally, since we threshold based on the similarity score in the retrieval stage, it is important to have comparable similarity scores across queries. Depending on the metric, some sounds may be inherently more similar to all other sounds, thus, we employ similarity normalization \cite{pizzi2022self, douze2021}.  For each query $q$, we %
discount its similarity with each reference $r$ using
a bias term based on the average similarity between $q$ and its $K$ nearest neighbors in a background set of other samples, resulting in the normalized similarity score:  %
\begin{equation}
\label{eq:score_norm}
    \begin{gathered}
    \hat{s}(q, r) = {s}(q, r) - \beta \cdot \text{bias}(q), \\
     \text{bias}(q) = \frac{1}{K} \sum_{k = 1}^{K} s(q, b_k),
    \end{gathered}
\end{equation}
where $b_k$ is the %
$k$-th nearest neighbour of $q$ in the background set (based on the similarity of their descriptors), %
and $\beta$ a scalar. Therefore, in the retrieval stage, we retrieve the following set of queries:
\begin{equation}
\label{eq:retrieval}
    \{q : \max_{r}{ \hat{s}({q}, {r})} \geq \tau \},
\end{equation}
which are then inspected along with their top-1 matches in the verification stage, based on the definition of Section~\ref{sec:intro}. 

\section{Experimental Framework}
\label{sec:exp}

The model we focus on and experiment with is Tango~\cite{ghosal2023tango}, firstly because its code (including training scripts) and model checkpoints are available online. Secondly, it maintains 
near state-of-the-art performance while being trained on AudioCaps~\cite{kim2019audiocaps}, which is a significantly smaller training set compared to similar models, containing around 45k audio and text pairs, thus making the search for replications more tractable. We note however that, despite the small training set size, level-weighted mixing augmentation of two sounds with concatenated captions is performed during training. Every sample generation is performed using $200$ diffusion steps and classifier-free guidance~\cite{ho2022classifier} with a scale of $3$. 

All sounds have a sampling frequency of \qty{16}{kHz} and are zero-padded to \qty{10.242}{s} to match the length of the generated samples. The mel %
spectrogram is computed using 16 mel bins and a \qty{128}{ms} long Hann window with $25\%$ overlap. It is then normalized by dividing it by its maximum value, converted to decibel, and clipped with a lower value of \qty{-40}{dB}. By flattening, we get a 1712-dimensional descriptor. In order to compute CLAP \cite{wu2023large} descriptors, we use the publicly available $\texttt{laion/clap-htsat-unfused}$ model checkpoint on Hugging Face, resulting in 512-dimensional descriptors. For the calculation of the normalized similarity score, we use a background set consisting of 1000 samples randomly selected from the balanced train segment of the AudioSet dataset \cite{gemmeke2017audio}, which has no overlap with AudioCaps. We set $\beta = 0.5$ and use the top-5 nearest neighbors from the background set, i.e., $K = 5$.

\section{Results \& Discussion}
\label{sec:results}

\subsection{Comparison of Descriptors}

First, we compare the mel and CLAP descriptors in terms of detecting replicated samples. To that end, we train two additional Tango models on small subsets of the AudioCaps training set, while still performing the mixing augmentation. The first model was trained on 1000 audio and text pairs, and the second on 5000. Each model is trained for $62200$ steps, the maximum number of training steps of the pre-trained checkpoint. With each model, we generate using the same training set prompts, resulting in as many generated samples as the size of each training set. We observe extreme overfitting, in terms of difference between the validation and training loss in the case of the first model, and slight overfitting in the case of the second. 

\begin{figure}[t!]
    \centering
    \includegraphics[width=\linewidth]{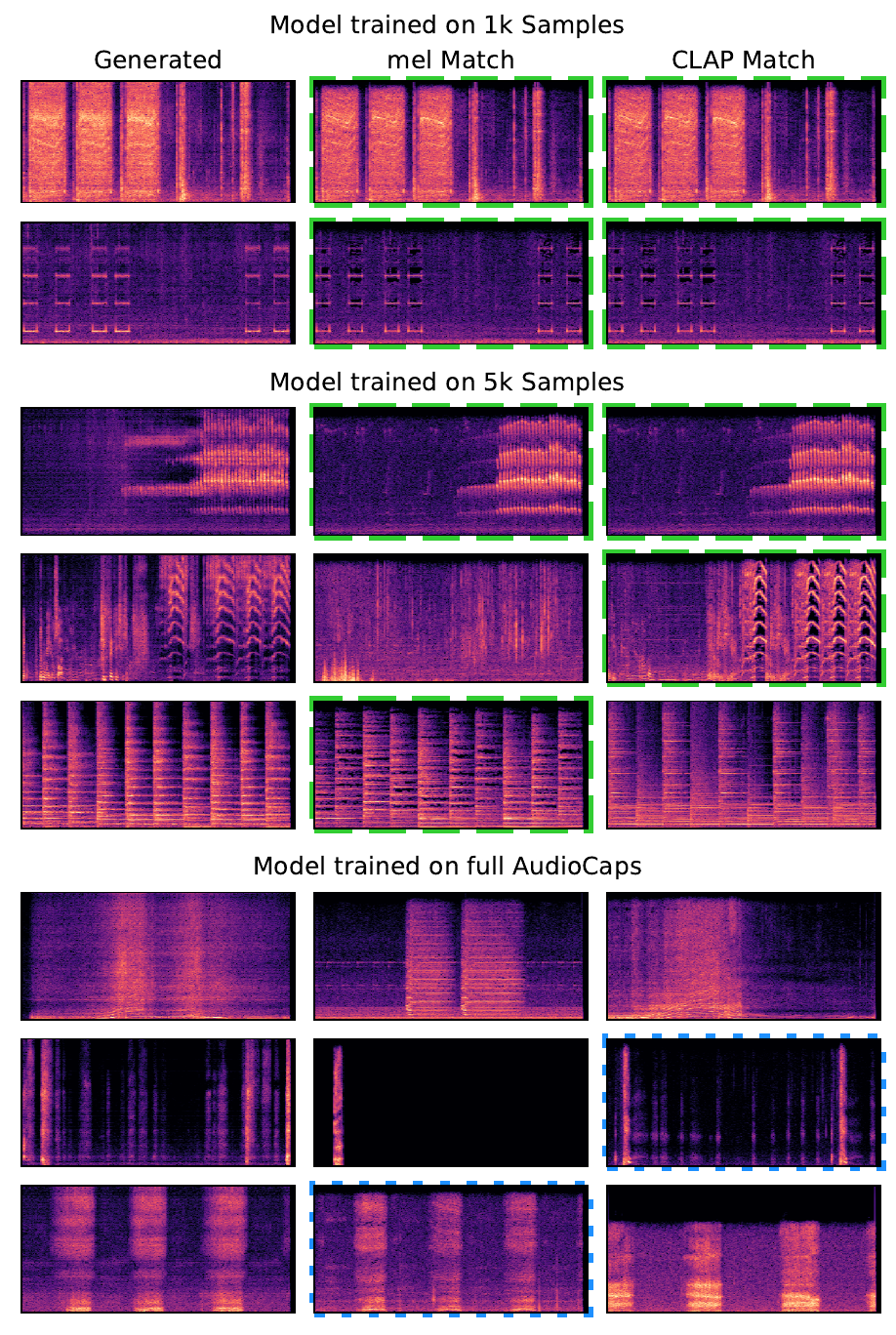}
      \caption{Spectrograms of generated samples (left column) along with their top-1 matches using mel descriptors (middle column) and CLAP descriptors (right column) for each model. Green dashed outlines indicate successful match detection, while the blue dotted outlines indicate detection of a potential match. }
      \label{fig:mel_clap_matches}
\end{figure}

\begin{figure}[t!]
    \centering
    \includegraphics[width=\linewidth]{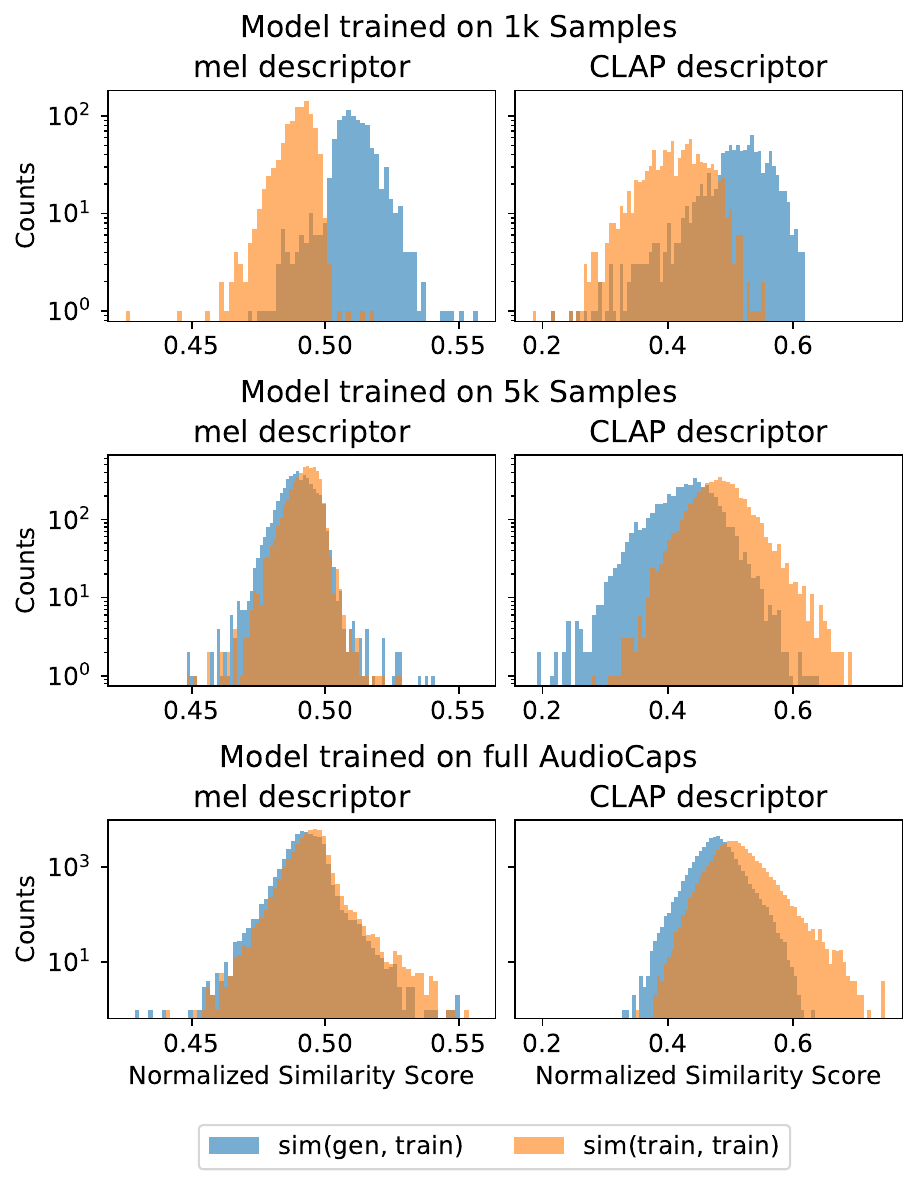}
      \caption{Histograms of the top-1 normalized similarity scores, using both descriptors, between the generated samples and the corresponding training samples (blue), as well as between the training samples and themselves (orange). The vertical axis is in logarithmic scale.}
      \label{fig:mel_clap_hist}
\end{figure}

As expected, we observe widespread replication in the generations of the first model, with over $90$\% of them being almost identical replications of training samples. Two examples of memorized generations are shown in the top of Fig.~\ref{fig:mel_clap_matches}, together with their matches successfully identified by both mel and CLAP methods.
Memorization can also be seen in the first row of Fig.~\ref{fig:mel_clap_hist}, where we observe clear separation of the distributions of the top-1 normalized similarity scores of the generations to the training set, and of the training set to itself. We note that the mel method achieves better separation, which can translate into better precision in replication detection. Based on these histograms, we select the threshold value $\tau_\text{mel} = 0.5005$ for the following analyses of the two other models.  

For the second model trained on 5000 training samples, we follow the methodology outlined in Section~\ref{sec:method} using mel and CLAP descriptors. We set $\tau_\text{mel} = 0.5005$, resulting in 178 pairs, and we appropriately set $\tau_\text{CLAP}$ to get the same number of pairs. Out of those, following manual verification using the definition from Section~\ref{sec:intro}, just 31 out of 178 are replicated in the case of mel descriptors and 28 out of 178 in the case of CLAP, with only 2 being in both sets.

We find that mel descriptors tend to pay attention to similar spectro-temporal energy presence, while CLAP descriptors focus mostly on similar semantic content. In Fig.~\ref{fig:mel_clap_matches} (middle), we see a failure case for each method. Also, in contrast to the previous model, replication pairs here are not always exact copies, but contain certain complex spectro-temporal patterns that are nearly identical. Finally, Fig.~\ref{fig:mel_clap_hist} (middle) confirms that replication is limited in comparison to the previous case, with the mel histograms more closely corresponding to our empirical observations (i.e, limited presence of replicated samples).

\subsection{Investigation of Replication}

We now shift our attention to the model trained on the whole AudioCaps dataset. As before, we generate around 45k samples using the text prompts of the training set and we then search for potential matches following the methodology described in Section~\ref{sec:method}. We use the threshold $\tau_\text{mel} = 0.5005$ that retrieves 2278 samples and the appropriate $\tau_\text{CLAP}$ that results in the same number of pairs. 
After manually inspecting those samples, the number of samples that potentially fit our replication definition is quite limited, around a dozen for each method with mel descriptors detecting slightly more. 

We show in Fig.~\ref{fig:mel_clap_matches} (bottom) a few generations and their closest mel and CLAP matches, demonstrating how both methods operate. In the first case, both methods do not return a match, with the mel method returning a sample with similar spectro-temporal distribution of energy and the CLAP method returning a semantically similar sample. In the second case, the CLAP method detects a training set sample containing a sneeze sound that temporally coincides with the first one in the generation, while the mel method returns a burp sound occurring at that same timestep. In the third and final case, we have a snoring sound, where the mel method finds a very close albeit not identical sample, while the CLAP method returns a semantically similar sample also consisting of snoring.

\begin{figure*}[t]
    \centering
    \includegraphics[width=\linewidth]{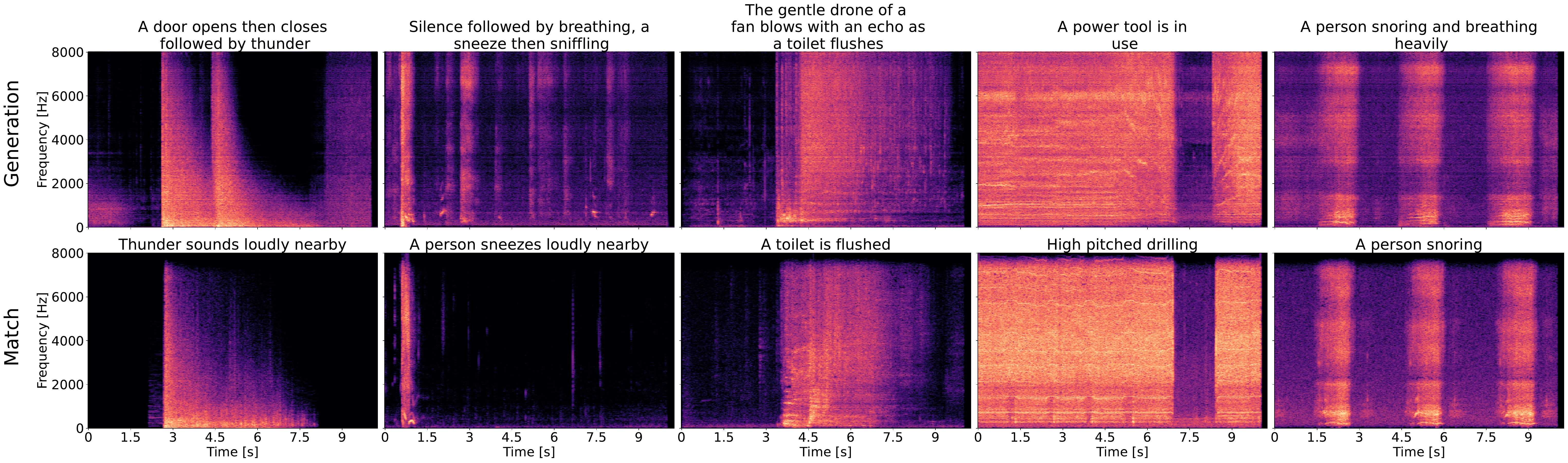}
    
    \includegraphics[width=\linewidth]{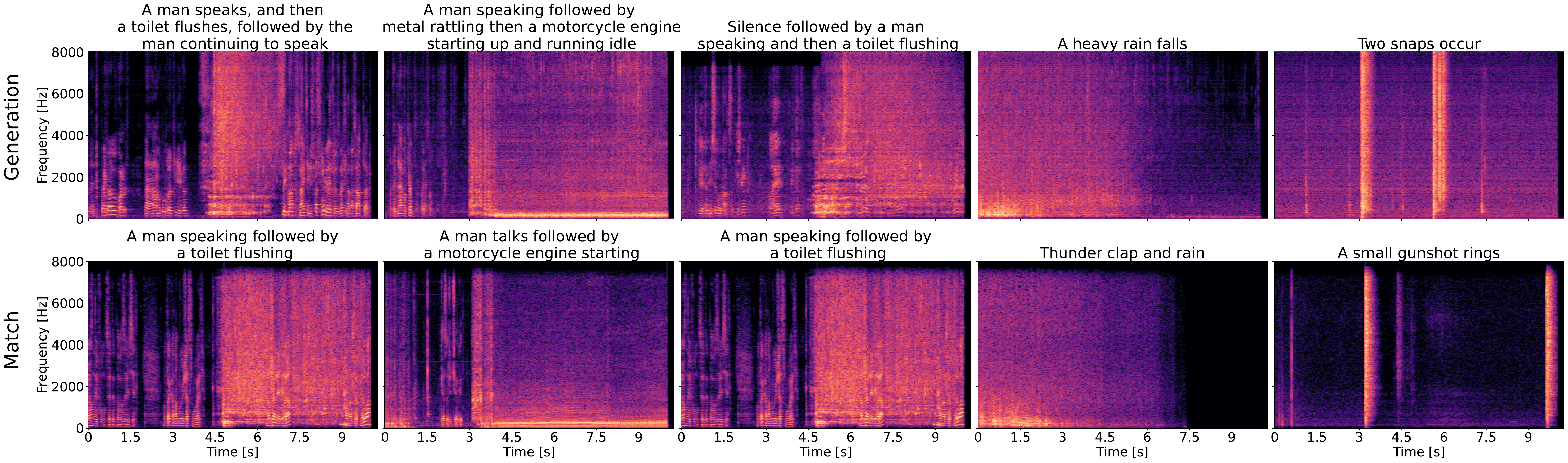}%
      \caption{Matches found using mel descriptors in the top two rows. Matches found using CLAP descriptors in the bottom two rows.}
      \label{fig:matches}
\end{figure*}

In Fig.~\ref{fig:matches}, we showcase some detected potential replication matches using the mel and CLAP methods. In the case of mel, we have very similar spectro-temporal patterns coinciding, especially in the temporal dimension. In the case of CLAP, we have semantic matches whose acoustic components are very similarly structured. Minor variations between the matches could be attributed to parts of the models (e.g., the vocoder) that are pre-trained on other datasets. Additionally, the presence of additional acoustic events in the generations might be a result of the augmentation used during training.

\subsection{Duplication in the AudioCaps Training Set}

The presence of training samples that form a tail of high self-similarity in the histogram in Fig.~\ref{fig:mel_clap_hist} (bottom row) prompts us to investigate duplication in the AudioCaps dataset. First, we compute the pairwise normalized self-similarity matrix $S$ between the audio data of the AudioCaps training set, using mel descriptors. Second, we zero out the diagonal of $S$. Next, we compute the binary matrix $A_{ij} = \mathds{1}\left[ S_{ij} > \tau \text{ and } S_{ji} > \tau\right]$ where $\tau = 0.5025$ and $\mathds{1}\left[ \cdot \right]$ represents the indicator function. The resulting symmetric matrix $A$ can be viewed as a graph adjacency matrix and we can find its connected components. Finally, we manually examine each found connected component and select ones that correspond to clusters of duplicated samples in the training set.

\begin{figure}[t!]
    \centering
    \includegraphics[width=\linewidth]{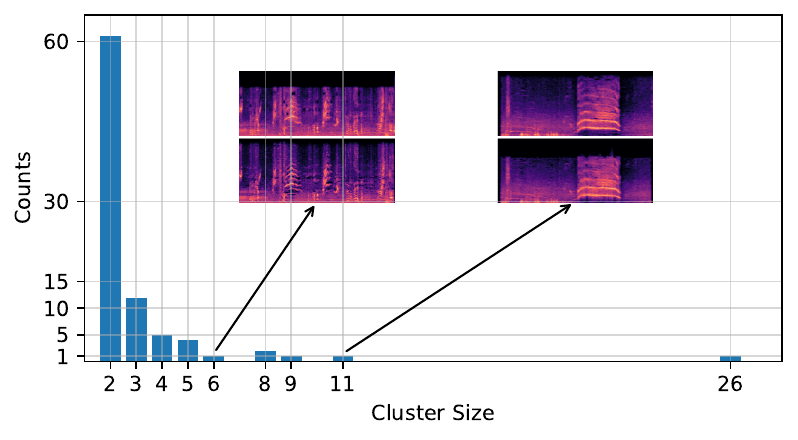}
      \caption{Clusters of audio duplicates found in the AudioCaps training set.}
      \label{fig:duplicated_clusters}
\end{figure}

We find 276 connected components where 88 of them are indeed duplicated clusters containing 257 samples in total. We present these findings in Fig.~\ref{fig:duplicated_clusters}. 
Examples include audio files extracted from different YouTube videos containing the same intro audio effects. 
However, despite the presence of duplicate audio files in the training set, we did not find generations replicating them.

\section{Conclusions and Future Work}
\label{sec:conclusion}

We performed an initial analysis of training data replication in \textsc{Tango}, a recently proposed state-of-the-art text-to-audio latent diffusion model. While we only manually verified a small number of replications on the model trained on the full dataset (approximately 5-10 per every 10,000 training samples), we have provided some evidence of training data memorization. We believe studies such as this one are important to help technically inform many of the outstanding legal issues related to generative modeling. While we compared mel and CLAP descriptors for finding replications, there are countless additional techniques to be explored, e.g., mel+CLAP ensembles, audio fingerprinting~\cite{cano2002review, chang2021neural}, self-supervised learning~\cite{gong2022ssast}, etc. We also surprisingly found duplicate sound files in AudioCaps, a curated subset of AudioSet. It has recently been observed that de-duplicated training data can improve performance in language models~\cite{kandpal2022deduplicating, lee2022deduplicating}, and a similar analysis for audio is an interesting area for further study.

\clearpage
\balance
\bibliographystyle{IEEEtran}
\bibliography{refs}

\end{document}